\documentclass[twocolumn, twocolappendix]{aastex701}

\usepackage{threeparttable}
\usepackage{multirow}
\usepackage{booktabs}
\usepackage{subfigure}
\usepackage{amsmath,amssymb,amsfonts}
\usepackage{amsthm}
\usepackage{CJK}

\newcommand{\src}{V4641\,Sgr}

\begin{document}
\begin{CJK*}{UTF8}{gbsn}
\title{Candidate X-ray Counterparts of Ultra-High-Energy Emission of the Galactic Microquasar \\ V4641 Sgr Revealed by Einstein Probe}

\author[0000-0002-2421-173X]{Yi-Heng Chi (池奕恒)}
\affiliation{School of Astronomy \& Space Science, Nanjing University, 163 Xianlin Avenue, Nanjing 210023, China}
\affiliation{Key Laboratory of Modern Astronomy and Astrophysics, Nanjing University, Ministry of Education, Nanjing 210023, China}
\email{yhchi@smail.nju.edu.cn}

\author{Shao-Qiang Xi (席绍强)}
\affiliation{State Key Laboratory of Particle Astrophysics, Institute of High Energy Physics, Chinese Academy of Sciences, Beijing 100049, China}
\affiliation{Tianfu Cosmic Ray Research Center, 610000 Chengdu, Sichuan, China}
\email[show]{xisq@ihep.ac.cn}

\author[0000-0002-5683-822X]{Ping Zhou (周平)}
\affiliation{School of Astronomy \& Space Science, Nanjing University, 163 Xianlin Avenue, Nanjing 210023, China}
\affiliation{Key Laboratory of Modern Astronomy and Astrophysics, Nanjing University, Ministry of Education, Nanjing 210023, China}
\email[show]{pingzhou@nju.edu.cn}

\author[0000-0001-7584-6236]{Hua Feng (冯骅)}
\affiliation{State Key Laboratory of Particle Astrophysics, Institute of High Energy Physics, Chinese Academy of Sciences, Beijing 100049, China}
\email[show]{hfeng@ihep.ac.cn}

\author[0000-0003-1576-0961]{Ruo-Yu Liu (柳若愚)}
\affiliation{School of Astronomy \& Space Science, Nanjing University, 163 Xianlin Avenue, Nanjing 210023, China}
\affiliation{Key Laboratory of Modern Astronomy and Astrophysics, Nanjing University, Ministry of Education, Nanjing 210023, China}
\affiliation{Tianfu Cosmic Ray Research Center, 610000 Chengdu, Sichuan, China}
\email[show]{\\ryliu@nju.edu.cn}

\author[0000-0003-0703-1275]{Songzhan Chen (陈松战)}
\affiliation{State Key Laboratory of Particle Astrophysics, Institute of High Energy Physics, Chinese Academy of Sciences, Beijing 100049, China}
\affiliation{Tianfu Cosmic Ray Research Center, 610000 Chengdu, Sichuan, China}
\email{szchen@ihep.ac.cn}

\author[0000-0001-7209-9204]{Naomi Tsuji}
\affiliation{Institute for Cosmic Ray Research, The University of Tokyo, 5-1-5, Kashiwa-no-ha, Kashiwa, Chiba 277-8582, Japan}
\affiliation{Interdisciplinary Theoretical \& Mathematical Science Center (iTHEMS), RIKEN, 2-1, Hirosawa, Wako, Saitama 351-0198, Japan}
\email{ntsuji@icrr.u-tokyo.ac.jp}

\author[0009-0002-1992-0707]{Caterina Tresoldi}
\affiliation{INAF-Osservatorio Astronomico di Brera, Via E. Bianchi 46, 23807 Merate (LC), Italy}
\affiliation{Como Lake Center for Astrophysics (CLAP), DiSAT, Universitàdegli Studi dell’Insubria, via Valleggio 11, 22100 Como, Italy}
\email{caterina.tresoldi@inaf.it}

\author[0000-0003-0293-3608]{Gabriele Ponti}
\affiliation{INAF-Osservatorio Astronomico di Brera, Via E. Bianchi 46, 23807 Merate (LC), Italy}
\affiliation{Como Lake Center for Astrophysics (CLAP), DiSAT, Universitàdegli Studi dell’Insubria, via Valleggio 11, 22100 Como, Italy}
\email{gabriele.ponti@inaf.it}

\author[0000-0002-4858-6130]{Zhen Cao (曹臻)}
\affiliation{State Key Laboratory of Particle Astrophysics, Institute of High Energy Physics, Chinese Academy of Sciences, Beijing 100049, China}
\affiliation{University of Chinese Academy of Sciences, 100049 Beijing, China}
\affiliation{Tianfu Cosmic Ray Research Center, 610000 Chengdu, Sichuan, China}
\email{caozh@ihep.ac.cn}

\author[0000-0001-9834-2196]{Yong Chen (陈勇)}
\affiliation{State Key Laboratory of Particle Astrophysics, Institute of High Energy Physics, Chinese Academy of Sciences, Beijing 100049, China}
\email{ychen@ihep.ac.cn}

\author[0000-0002-5203-8321]{Shumei Jia (贾淑梅)}
\affiliation{State Key Laboratory of Particle Astrophysics, Institute of High Energy Physics, Chinese Academy of Sciences, Beijing 100049, China}
\email{jiasm@ihep.ac.cn}

\author[0000-0002-2006-1615]{Chi-Chuan Jin (金驰川)}
\affiliation{National Astronomical Observatories, Chinese Academy of Sciences, 20A Datun Road, Beijing 100101, China}
\affiliation{School of Astronomy and Space Science, University of Chinese Academy of Sciences, 19A Yuquan Road, Beijing 100049, China}
\email{ccjin@nao.cas.cn}

\author[0000-0001-5798-4491]{Chengkui Li (李承奎)}
\affiliation{State Key Laboratory of Particle Astrophysics, Institute of High Energy Physics, Chinese Academy of Sciences, Beijing 100049, China}
\email{lick@ihep.ac.cn}

\author[0009-0007-6104-337X]{Yuan Liu (刘元)}
\affiliation{National Astronomical Observatories, Chinese Academy of Sciences, 20A Datun Road, Beijing 100101, China}
\email{liuyuan@bao.ac.cn}

\author[0000-0003-3802-1619]{Matthew Lundy}
\affiliation{Columbia Astrophysics Laboratory, Columbia University, 538 West 120th Street, New York, NY 10027, USA}
\email{mwl2145@columbia.edu}

\author[0000-0002-9709-5389]{Kaya Mori}
\affiliation{Columbia Astrophysics Laboratory, Columbia University, 538 West 120th Street, New York, NY 10027, USA}
\email{kaya@astro.columbia.edu}

\author[0000-0001-5249-2797]{Haiwu Pan (潘海武)}
\affiliation{National Astronomical Observatories, Chinese Academy of Sciences, 20A Datun Road, Beijing 100101, China}
\email{panhaiwu@bao.ac.cn}

\author[0000-0002-2705-4338]{Lian Tao (陶炼)}
\affiliation{State Key Laboratory of Particle Astrophysics, Institute of High Energy Physics, Chinese Academy of Sciences, Beijing 100049, China}
\email{taolian@ihep.ac.cn}

\author[0000-0003-4874-0369]{Junfeng Wang (王俊峰)}
\affiliation{Department of Astronomy, Xiamen University, Xiamen 361005, China}
\email{jfwang@xmu.edu.cn}

\author[0000-0001-8266-3024]{Weimin Yuan (袁为民)}
\affiliation{National Astronomical Observatories, Chinese Academy of Sciences, 20A Datun Road, Beijing 100101, China}
\affiliation{School of Astronomy and Space Science, University of Chinese Academy of Sciences, 19A Yuquan Road, Beijing 100049, China}
\email{wmy@nao.cas.cn}

\author[0000-0001-8869-0672]{Juan Zhang (张娟)}
\affiliation{State Key Laboratory of Particle Astrophysics, Institute of High Energy Physics, Chinese Academy of Sciences, Beijing 100049, China}
\email{zhangjuan@ihep.ac.cn}

\author[0000-0002-2192-1461]{Haisheng Zhao (赵海升)}
\affiliation{State Key Laboratory of Particle Astrophysics, Institute of High Energy Physics, Chinese Academy of Sciences, Beijing 100049, China}
\email{zhaohs@ihep.ac.cn}

\begin{abstract}

Microquasars have emerged as promising PeVatron candidates capable of accelerating particles to PeV energies, with recent observations revealing ultra-high-energy (UHE) $\gamma$-ray emission from extended structures associated with these systems. Multi-wavelength studies of such extended structures are crucial for understanding the acceleration, transport, and radiation processes of UHE cosmic rays (CRs), yet observational evidence remains limited. Here, we report the discovery of diffuse X-ray structures spatially coincident with the TeV $\gamma$-ray emission regions to the north and south of the microquasar V4641\,Sgr, using the Follow-up X-ray Telescope (FXT) onboard the Einstein Probe. We identify X-ray structures located $\sim20$--$40$ pc from V4641\,Sgr, extending over several arcminutes. These structures exhibit X-ray fluxes of $\sim1$--$3\times10^{-13}$\,erg\,s$^{-1}$\,cm$^{-2}$ and photon indices of $\Gamma\sim2$--3 in 0.5--8\,keV per individual knot.  
Under the one-zone leptonic interpretation, the weak X-ray emission implies a magnetic field of order $1-2\,\mu$G. Stronger fields remain possible if only part of the TeV--PeV emission is produced by inverse Compton scattering, with the remaining emission arising from hadronic interactions. The EP-FXT observations support the presence of relativistic electrons on tens-of-parsec scales, while the origin of the UHE $\gamma$-ray emission remains open.
\end{abstract}

\keywords{\uat{X-ray astronomy}{1810} --- \uat{Gamma-ray sources}{633} --- \uat{High Energy astrophysics}{739} --- \uat{Cosmic rays}{329} --- \uat{Jets}{870} --- \uat{Non-thermal radiation sources}{1119}}

\section{Introduction}
Cosmic rays (CRs) with energy lower than $\sim3\times10^{15}$\,eV ($\sim3$\,PeV) are generally believed to originate from Galactic sources. High-energy stellar objects including pulsar wind nebulae (PWNe) and supernova remnants (SNRs) prove to be dominant GeV-to-TeV CR accelerators for both their extensive associations with GeV-TeV $\gamma$-ray sources and satisfying the energy budget. However, the origin of the highest-energy Galactic CRs approaching several PeVs (so called ``PeVatrons'') remains an open question \citep[e.g.,][]{Cao21, Cao24}. Recently, LHAASO reported the detection of ultra-high-energy ($\gtrsim100$\,TeV; UHE) $\gamma$-ray emissions from several microquasars \citep{Cao25}. With $\gamma$-ray photon energy up to $\sim1$\,PeV and no evident high-energy cutoff in the spectrum, these black hole-jet systems emerge as new and promising PeVatron candidates.

Compared to SNRs and PWNe, multi-wavelength study of microquasars and their environment is still limited for understanding the acceleration, transport, and radiation of UHE CRs. One exception is SS\,433. Recent observations revealed that TeV $\gamma$-ray morphology \citep{HESS24} extends along the large-scale synchrotron X-ray jets \citep{Waston83, Yamauchi94, Safi-Harb22}, suggesting that particles may be accelerated at the ``jet base'' and then transported along the jet, with synchrotron cooling shaping the X-ray spectra \citep{HESS24}.  
On the (sub-)arcsecond scales, the relativistic jets have been resolved in radio \citep{Paragi99}, optical \citep{Waisberg19}, and X-ray band \citep{Sakai25}, revealing detailed jet structures.
On the large scales up to 100--200\,pc, shells and thermal diffuse emission have been detected in radio \citep[i.e., the Manatee Nebula][]{Dubner98} and X-rays \citep[e.g.,][]{Safi-Harb97, Chi24}. These multi-wavelength observations provide critical constraints on the origin  and radiation mechanisms of UHE $\gamma$-ray emissions in SS\,433 \citep{Cao25}. 

However, except for SS 433, the multi-wavelength properties of most UHE microquasars remain poorly constrained. Among them, V4641\,Sgr is particularly intriguing.
The microquasar V4641\,Sgr is a black hole X-ray binary \citep{Orosz01} known for its giant outburst in September 1999 \citep{Stubbings99, McCollough99}, accompanied by relativistic jets \citep{Hjellming00}. This system has recently drawn renewed attention for its extended UHE $\gamma$-ray emissions detected by LHAASO \citep{Cao25} and HAWC \citep{HAWC24}.  Its $\gamma$-ray flux above 100\,TeV is over one order of magnitude higher than those of other known microquasars, with a spectrum extending to 0.8\,PeV without evident spectral cutoff \citep{Cao25}. The high flux in TeV $\gamma$-rays make it an ideal target to search for multi-wavelength counterparts.

The morphology of the TeV $\gamma$-ray emission around V4641~Sgr is also intriguing. LHAASO \citep{Cao25}, HAWC \citep{HAWC24}, and H.E.S.S. \citep{HESS25} all found the source can be either described by two separate sources located at two opposite sites north and south of the binary, or by an elongated source centered close to the binary. The morphology may reflect the formation mechanism of the $\gamma$-ray source. In particular, the former scenario resemble that of SS\,433. Therefore, a view from its multiwavelength counterpart would be crucial to reveal the true nature of the $\gamma$-ray source.

In recent years, several efforts have been made to search for multi-wavelength emissions surrounding V4641\,Sgr. \citet{Noa26} discovered a faint radio bow-tie-shaped structure misaligned with the jets of V4641\,Sgr. In the X-ray, \citet{Suzuki25} reported an extended structure with an angular size of $\sim20$\,pc centered on the microquasar. However, the observation was performed during the outburst on September 30, 2024, and contamination due to the PSF (point-spread function) wings of the luminous point source or a potential scattering halo cannot be excluded. In addition, contamination from nearby point-like sources may be difficult to exclude due to the limited angular resolution of XRSIM/Xtend \citep[half power diameter, HPD, of $\sim1.7'$,][]{Mori24}. 

Therefore, follow-up X-ray observations are essential for constraining the spectral and spatial properties of the X-ray structures near V4641\,Sgr. The Follow-up X-ray Telescope \citep[FXT,][]{Chen20} onboard the Einstein Probe \citep[EP,][]{Yuan22} has a powerful capability to search for faint extended X-ray emissions thanks to its large field of view ($\sim1^\circ \times \sim1^\circ$) and low instrumental background \citep{Zhang25}. Meanwhile, the Wide-field X-ray Telescope (WXT) onboard EP, based on lobster-eye optics, enables long-term monitoring of V4641\,Sgr, which is crucial for identifying if there were strong X-ray variability during the FXT observational epochs.

In this letter, we present our FXT study of V4641\,Sgr, coordinated with a contemporaneous XMM-Newton study (Tsuji et al.). We describe our observations in \S\ref{sec: xobs} and report the detection of diffuse emissions north and south of V4641\,Sgr in \S\ref{sec:res}. We further compare V4641\,Sgr and SS\,433 in terms of X-ray and $\gamma$-ray morphology in \S\ref{sec:xray}. A broadband spectral distribution (SED) from X-ray to $\gamma$-ray is presented in \S\ref{sec:gam}, and we discuss the magnetic field and leptonic emissions around V4641\,Sgr. The paper is summarized in \S\ref{sec:summary}.

\section{EP-FXT observations}\label{sec: xobs}
The V4641\,Sgr region was observed by EP-FXT between 2025 and 2026 in the Full Frame mode with the Thin filter. These observations were pointed to the northern and southern TeV $\gamma$-ray sources, covering a sky area of $1.6^\circ\times1.1^\circ$ and accumulating a total exposure time of $\sim64$\,ks. A long-term monitor was carried out by EP-WXT, where the flux of V4641\,Sgr is lower than $5\times10^{-11}$\,erg\,s$^{-1}$\,cm$^{-2}$ so that its contamination (e.g., PSF patterns and stray light) to the diffuse emissions is negligible. The detailed observation information is listed in Appendix~\ref{app: obs}. 

We used the FXT Data Analysis System (FXTDAS, ver. 1.30 with FXT CALDB ver.\ 1.30) to process the data, with some analysis made using HEASoft \citep[ver. 6.36,][]{heasoft14} and CIAO \citep[ver. 4.18,][]{Fruscione06} software. The observations were recalibrated, filtered, and reprojected as described in \citet{Cao26}. The instrumental background was generated by {\it fxtbkggen} according to the events in the frame storage area, and then rescaled according to the 9--11-keV X-ray counts to match the level of observed images and spectra. The background-subtracted images were divided by the exposure map generated by {\it fxtexpogen} to correct the vignetting. CIAO script {\it wavdetect} was used to search for point-like sources, which are re-filled with {\it roi} and {\it dmfilth} to underline diffuse emissions. The images were adaptively smoothed by {\it dmimgadapt}.

We extracted spectra from each observation with {\it Xselect} and generated the response files using {\it fxtarfgen} and {\it fxtrmfgen}. All the spectra from the same instrument (FXTA or FXTB) and region are combined with {\it ftaddspec}. Finally, the combined spectra are grouped with a minimum of 20 counts per bin to apply the chi-square ($\chi^2$) statistics. The spectral analysis was applied with XSPEC \citep[][ver.\ 12.15.1]{Arnaud96} based on AtomDB \citep[][ver.\ 3.1.3]{Foster12}.

\begin{figure*}
    \centering
    \includegraphics[width=\textwidth]{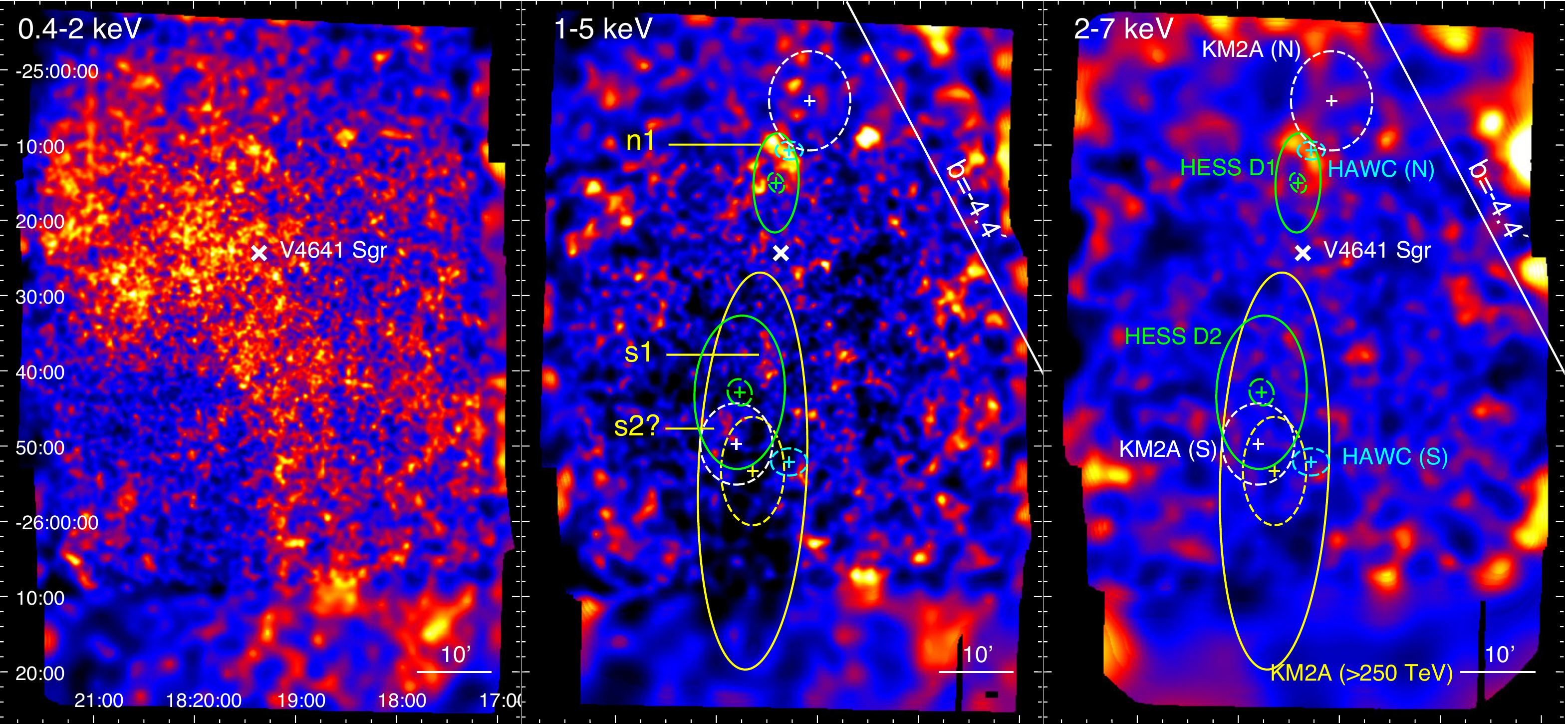}
    \caption{Background subtracted, vignetting corrected, and adaptive smoothed FXT images in 0.4--2 (left), 1--5 (middle), and 2--7 (right) keV with point-like sources removed. Diffuse X-ray structures n1, s1, and s2 are labeled in the middle panel. HAWC \citep[$>1$\,TeV]{HAWC24}, LHAASO/KM2A \citep[$>25$\,TeV, two point sources scenario therein]{Cao25}, ultra-high-energy KM2A \citep[$>250$\,TeV]{Cao25} and H.E.S.S. \citep[0.8--22\,TeV, Model D therein, which is also used for extracting spectra of D1 and D2]{HESS25} sources are marked in cyan, white, yellow, and green, respectively. Crosses (``+'') label the source center, dashed lines show the position error of the center, and solid lines are the 1$\sigma$ extension. The position of V4641\,Sgr is marked with a white ``X'' label.}
    \label{fig: x+gam}
\end{figure*}

\begin{figure*}
    \centering
    \includegraphics[width=\textwidth]{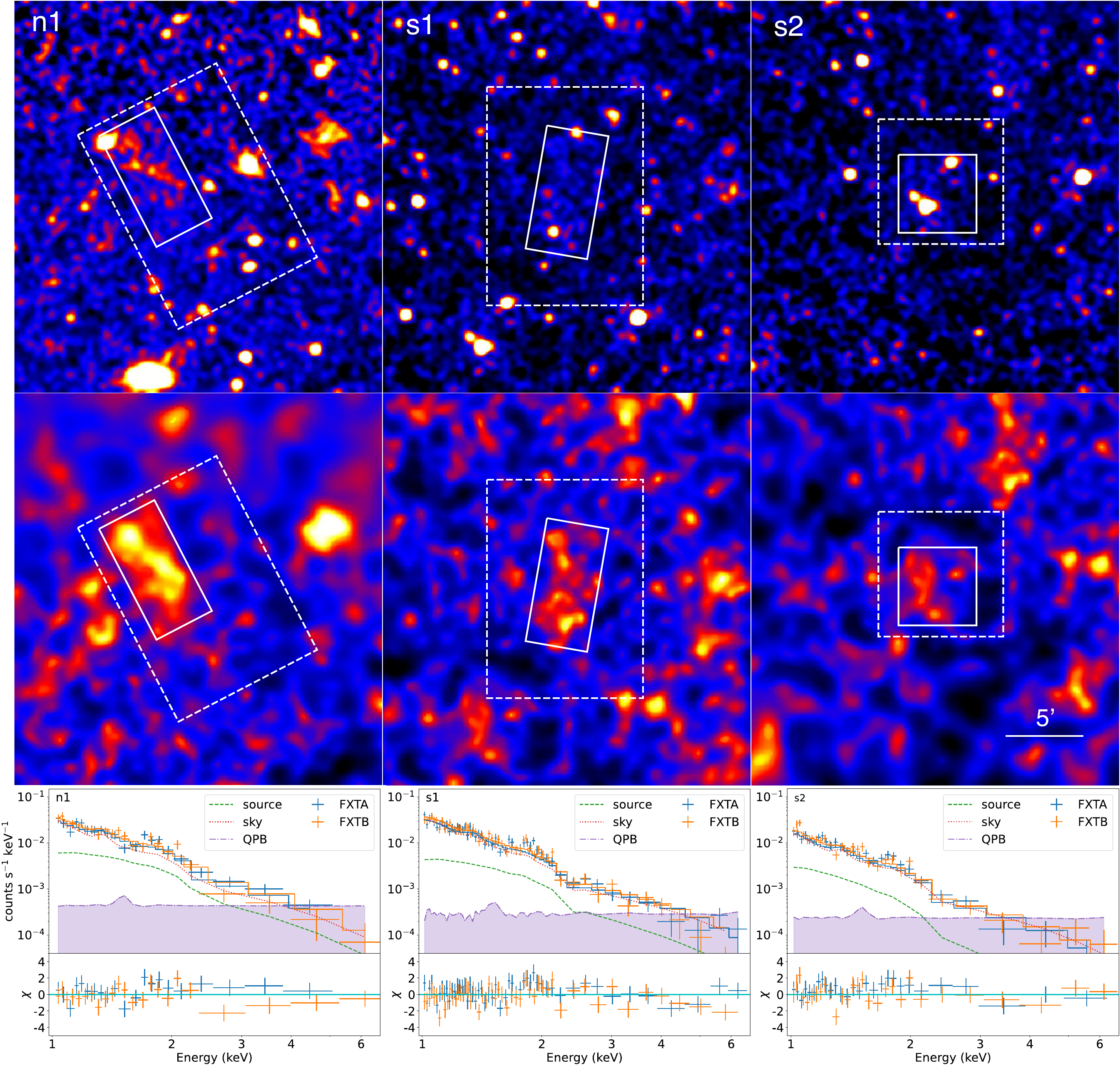}
    \caption{Zoomed-in vignetting corrected 1--5\,keV images (top and middle rows) and spectra with folded models (bottom panels) of n1 (left), s1 (middle), and s2 (right). The top panels are smoothed with a radius of $24''$ with a sampling rate of $4''$ per pixel and point-like sources maintained. The middle panels are point-like sources removed and adaptively smoothed as in Figure~\ref{fig: x+gam}. The source and background regions are marked with solid and dashed lines, respectively. For the spectra in the bottom panels, the purple lines mark the instrumental background level, which is subtracted from the observed spectra. Dashed green lines and dotted red lines stand for the source emission and the pre-fitted sky background, respectively.}
    \label{fig: reg}
\end{figure*}

\section{Results}\label{sec:res}

Figure~\ref{fig: x+gam} shows the 0.4--2\,keV, 1--5\,keV and 2--7\,keV FXT images of V4641\,Sgr and its vicinity, overlaid with the positions and extensions of $\gamma$-ray sources. Point-like sources have been removed from these images.
The 1--5\,keV energy band is an optimal selection to search for nonthermal emission, while thermal emission of diffuse X-ray background may dominate soft X-ray emission (also see the spectra in Appendix~\ref{app: bkg}). Moreover, FXT's effective area peaks at the 1--2\,keV band, and instrumental particle background dominates beyond 5\,keV.

FXT discovered several diffuse emissions, of which we especially focus on those with good spatial correlation with $\gamma$-ray sources as shown in Figures~\ref{fig: x+gam} and \ref{fig: reg}. The most prominent feature is a belt-like structure north of \src, named ``n1'' hereafter, with a size of $\sim4'\times8'$ (see the middle panel of Figure~\ref{fig: reg}), or $\sim7.2\,\mathrm{pc}\times14.4\,\mathrm{pc}$ at a distance of 6.2\,kpc \citep[see][and references therein]{HESS25}. It shows a good spatial correlation with the northern $\gamma$-ray source: it is overlapped by $1\sigma$ extension of the H.E.S.S. source D1 and correlated with the HAWC and LHAASO source within $\sim1\sigma$ uncertainties. 

In the south, the 1--5 keV image reveals some diffuse structures, of which the most promising one is ``s1''. Its extension is also $\sim4'\times8'$, while compared to n1, s1 has a lower peak surface brightness and some knot-like sub-structures. Although overlapped by the extension of H.E.S.S. source D2, s1 appears to the north of the southern source of HAWC and LHAASO. In the southern part of the H.E.S.S. D2, a diffuse residual appears after removal of point-like sources. We named it ``s2'', which is fainter than s1 and has an extension of $\sim5'$. It should be noted that s1 and s2 are local enhancement spatially overlapped with $\gamma$-ray sources, while there are several other likely diffuse emissions in the FoV (see Figure~\ref{fig: x+gam}). Also, we cannot firmly preclude the possibility that s1 and s2 are unrelated background sources, for example, associated with galaxy groups or clusters. Near the southernmost part of the LHAASO extension above 250\,TeV, there are no signatures of diffuse X-ray emissions. 

In the soft energy band (0.4--2 keV), a northeast-to-southwest enhancement is observed across the entire FXT FoV, which is likely associated with the foreground hot ionized gas distribution (also see Appendix~\ref{app: obs} for soft X-ray emissions in a larger sky area). However, we do not find evident thermal X-ray structures associated with V4641\,Sgr, such as the jet-heated gas \citep{Safi-Harb97} or large-scale shells \citep{Chi24} observed around SS\,433.

We further analyzed the spectra of n1, s1, and s2 as labeled in Figure~\ref{fig: reg} with the solar abundance according to \citet{Wilms2000}. We subtracted the instrumental background from the spectra of the source and the nearby sky regions, which are shown in Figure~\ref{fig: reg}, respectively. The off-source sky background was modeled (see details in Appendix~\ref{app: bkg}) 
and included as a background component in the spectral fitting of the source regions (Figure~\ref{fig: reg}).
The foreground absorption is difficult to constrain well due to the limited statistics of the faint X-ray structures. Therefore, we fixed the foreground column density to $2\times10^{21}$\,cm$^{-2}$ according to the HI values adopted from HI4PI \citep[][$1.7\times10^{21}$\,cm$^{-2}$]{HI4PI16} and previous studies of V4641\,Sgr \citep[][$2.5\times10^{21}$\,cm$^{-2}$]{Shaw22}.

\renewcommand\arraystretch{1.2}
\setlength{\tabcolsep}{2pt}
\begin{table*}
    \centering
    \caption{Fitting results with $1\sigma$ uncertainties. }
    \begin{tabular}{cccccccc}
    \hline
        Source & R.A. & Decl. & $\Gamma$ & $F_{0.5-8}^b$ & $F_{5\sigma}^c$ & $\Sigma_{0.5-8}^d$ & $\chi^2$/dof$^e$ \\ 
        & (J2000d) & (J2000d) & & ($10^{-13}$\,erg\,s$^{-1}$\,cm$^{-2}$) & ($10^{-13}$\,erg\,s$^{-1}$\,cm$^{-2}$) & ($10^{-15}$\,erg\,s$^{-1}$\,cm$^{-2}$\,arcmin$^{-2}$) & \\ \hline
        n1 & 274.83 & $-25.19$ & $1.71_{-0.33}^{+0.37}$ & $2.42\pm0.45$ & 0.82 & $7.6\pm1.4$ & 46.30/42 \\
        s1 & 274.87 & $-25.64$ & $2.09_{-0.26}^{+0.31}$ & $1.17\pm0.14$ & 0.52 & $3.7\pm0.4$ & 116.57/102 \\
        s2 & 274.96 & $-25.79$ & $3.02_{-0.50}^{+0.58}$ & $0.89_{-0.14}^{+0.18}$ & 0.30 & $3.6_{-0.6}^{+0.7}$ & 59.97/55 \\ \hline
        D1$^f$ & $\cdots$ & $\cdots$ & $2.09_{-0.12}^{+0.13}$ & $10.0(6.4)\pm0.5$ & $\cdots$ & $16.1(10.3)\pm0.8$ & 164.50/157 \\
        D2$^f$ & $\cdots$ & $\cdots$ & $1.73_{-0.08}^{+0.08}$ & $21.4(10.3)\pm0.5$ & $\cdots$ & $11.1(5.4)\pm0.3$ & 609.10/488 \\ \hline
    \end{tabular}
    \begin{tablenotes}
    \footnotesize
        \item $^a$Centroid J2000 coordinate of the adaptive smoothed image in the 1--5\,keV of source regions (e.g., lower panels of Figure~\ref{fig: x+gam}).
        \item $^b$Absorption corrected integrated flux in 0.5--8\,keV given by the XSPEC model {\it cflux}. The values in the brackets of D1 and D2 represent the flux after subtraction of the expected extragalactic background level.
        \item $^c$5-$\sigma$ lower boundary of $^b$.
        \item $^d$Surface brightness in 0.5--8\,keV.
        \item $^e$\,$\chi^2$: chi-square statistics; dof: degrees of freedom.
        \item $^f$ The regions are adopted from \citet{HESS25}.
    \end{tablenotes}
    \label{tab: xspec}
\end{table*}

The source emission was reproduced by a {\it powerlaw} model absorbed by {\it TBabs} \citep{Wilms2000}. The fitting results with 1$\sigma$ confidence level are shown in Table~\ref{tab: xspec} and the spectra with folded models are plotted in the bottom panels of Figure~\ref{fig: reg}. For regions n1, s1, and s2, the absorption corrected integrated fluxes in 0.5--8\,keV (the same below, except otherwise specified) are 2.42, 1.17, and $0.89\times10^{-13} \,\rm erg\,s^{-1}\,cm^{-2}$, respectively, with 1$\sigma$ uncertainties of 10--20\%. The corresponding photon indices are $\Gamma=1.7^{+0.4}_{-0.3}$, $2.1\pm 0.3$, $3.0^{+0.6}_{-0.5}$, respectively. Although the photon index cannot be well constrained due to the limited statistics, the best-fit value of $\Gamma\lesssim3$ of three regions is consistent with, but does not uniquely establish, a nonthermal origin. We also calculate the 5-$\sigma$ lower flux limits of 8.2, 5.2, $3.0\times10^{-14}\,\rm erg\,s^{-1}\,cm^{-2}$ for n1, s1, and s2, respectively. Thus, all these regions show significant flux excess above their local background, as also shown in Figure~\ref{fig: reg}. 

A thermal plasma model {\it apec} was also tested in the spectral fittings, leading to a high temperature, of which the 90\% lower limits are 3.2, 2.7, and 1.5\,keV for n1, s1, and s2, respectively. The thermal model cannot be statistically distinguished from the {\it powerlaw} model. On the other hand the fitted temperature is significantly higher than that of the hot gas associated with SS\,433 \citep[$\sim0.2$--1\,keV][]{Safi-Harb97, Chi24}. Located tens of parsecs from the microquasar, it is unlikely to maintain a high temperature due to, for instance, adiabatic cooling.

We also extracted the spectra from the H.E.S.S. source D1 and D2 \citep[][also labeled with green elliptical regions in Figure~\ref{fig: x+gam}]{HESS25}, covering the emission of n1 and s1+s2, respectively. Without an off-source background, we only subtracted the instrumental background and then modeled the spectra following the same modeling procedure used for the sky background (see Appendix~\ref{app: bkg}). The fitting results and fitted spectra are presented in Table~\ref{tab: xspec} and Figure~\ref{fig: dspec} in Appendix~\ref{app: bkg}. It should be treated as a ``conservative upper limit'' as the sky background was not subtracted. For example, the cosmic X-ray background (CXB) from unresolved active galaxy nuclei is expected to contribute $\sim3.6\times10^{-13}$ and $\sim1.1\times10^{-12}$\,erg\,s$^{-1}$\,cm$^{-2}$ for D1 and D2, respectively, according to the CXB model provided in XMM-ESAS Cookbook\footnote{https://heasarc.gsfc.nasa.gov/docs/xmm/esas/cookbook/xmm-esas.html}. Besides, the surface brightnesses of D1 and D2 exceed that of n1 with local background subtracted, suggesting that if the flux here is dominated by synchrotron emissions of V4641\,Sgr rather than the sky background, we should observe locally enhancement. The diffuse emission from the Galactic Ridge \citep[e.g.,][]{Valinia98} cannot be neglected near V4641\,Sgr as well. Therefore, the X-ray synchrotron flux associated with UHE $\gamma$-ray emissions would be much lower. As CXB has been well constrained and consistent among different telescopes \citep[e.g.,][]{DeLuca04, Hickox06}, we also list the flux and surface brightness of these two regions with CXB subtracted in Table~\ref{tab: xspec}. We also use these lower values, i.e., 6.4 and $10.3\times10^{-13}$\,erg\,s$^{-1}$\,cm$^{-2}$ for D1 and D2, respectively, in the discussion below.

We end this section with comparison to other X-ray studies of diffuse emissions associated with V4641\,Sgr. \citet{Suzuki25} reported the first X-ray search for the diffuse emissions around V4641\,Sgr, and claimed a detection which can be described with a 1-$\sigma$ radius of $\sim7'$ and a flux of $\sim5\times10^{-12}$\,erg\,s$^{-1}$\,cm$^{-2}$ in 2--10\,keV, corresponding to $\sim9\times10^{-12}$\,erg\,s$^{-1}$\,cm$^{-2}$ in 0.5--8\,keV assuming a photon index of 2. Thus, we can deduce the average surface brightness within the 1-$\sigma$ radius, containing $39\%$ of the total flux, would be $\sim2\times10^{-14}$\,erg\,s$^{-1}$\,cm$^{-2}$\,arcmin$^{-2}$, around 2.6 times that of n1. As n1 looks quite evident with an effective FXT exposure of $\lesssim20$\,ks, the detection of the XRISM source would be expected with our current data, but no diffuse emissions were observed during the on-axis FXT observations of V4641\,Sgr in its quiescent phase in 2025 (for details of the FXT observations see Appendix~\ref{app: obs}).

Meanwhile, independent XMM-Newton observations (Tsuji et al.) were carried out in September 2025, whose results are generally consistent with our study: n1 and s1 were detected in the same position, and there were no evident emissions around V4641\,Sgr. The flux given in our study is slightly lower, which could be subject to system uncertainties due to, for instance, instrumental calibrations and differences in data processing (i.e., on- and off-source regions, point-like sources detections, etc.).

\section{Discussion}
\subsection{Comparison with SS 433}\label{sec:xray}
\begin{figure}
    \centering
    \includegraphics[width=0.47\textwidth]{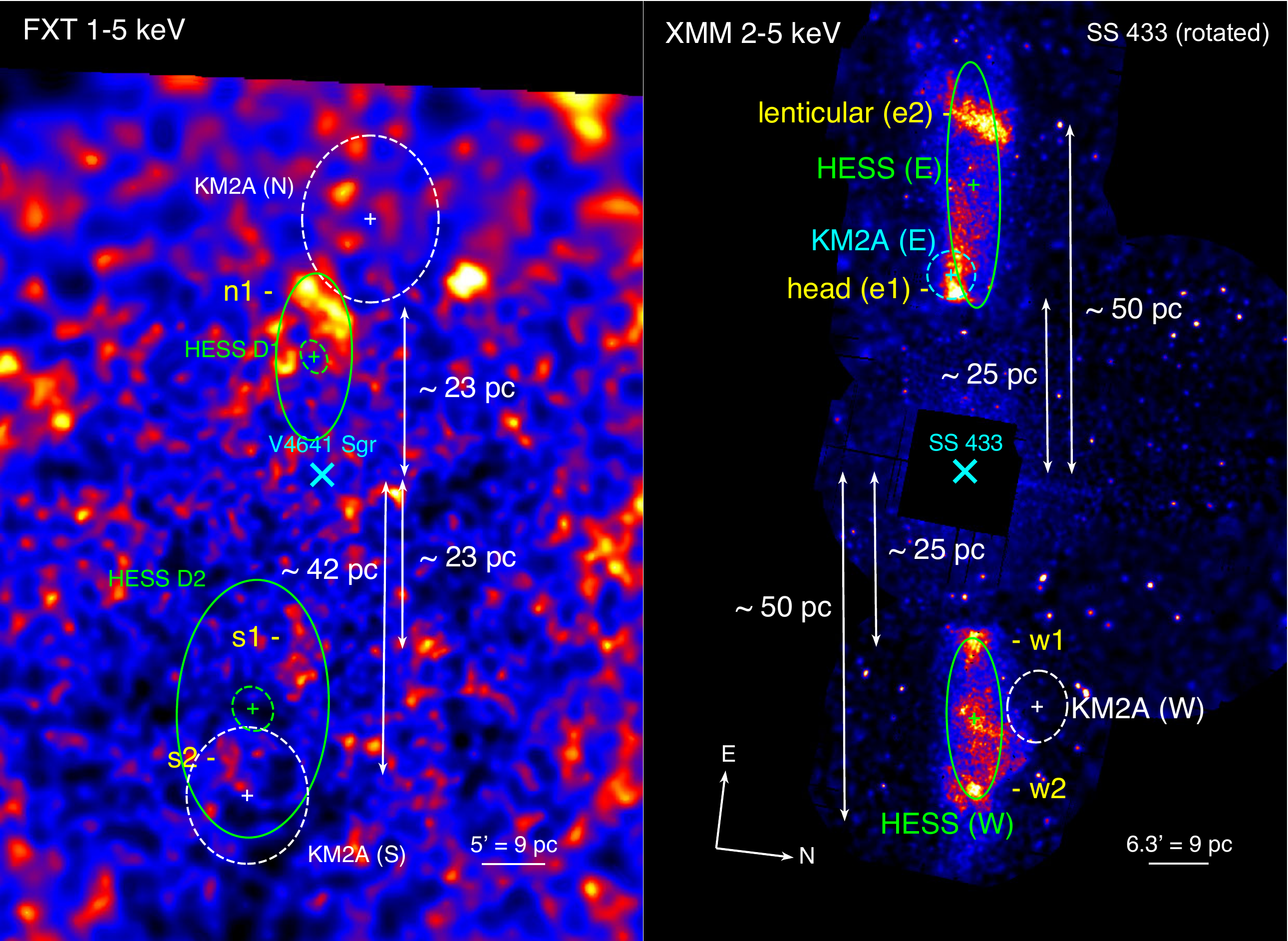}
    \caption{X-ray and $\gamma$-ray emissions associated with microqusars. From left to right: FXT 1--5\,keV image as in Figure~\ref{fig: x+gam}(a), XMM-Newton 2--5\,keV adaptively smoothed image derived from \citet{Chi24} of SS\,433 western and eastern jet. The images are zoomed to share the same physical scales. The H.E.S.S. position and extension are marked in green, and LHAASO/KM2A position and its uncertainties are marked with cyan crosses and dashed lines.}
    \label{fig: 4641+433}
\end{figure}

As the first microquasar exhibiting the best measurement of structures from radio to TeV $\gamma$-rays, SS\,433 serves as a representative source for investigating non-thermal emission surrounding microquasars. In the TeV band, two components modeled with elliptical Gaussian profiles are significantly detected to the east and west of SS\,433. The position and size of the TeV $\gamma$-ray components show spatial agreement with those of the X-ray emission \citep[][also see Figure~\ref{fig: 4641+433}]{Waston83, Safi-Harb22, HESS24}. If multi-wavelength emission surrounding SS\,433 shares a universal characteristic, analogous spatial structures should also be present in other microquasar systems. 

Among all TeV $\gamma$-ray sources associated with microquasars \citep{Cao25}, V4641\,Sgr stands out as the closest analog with its elongated $\gamma$-ray emission, which can be resolved into two elliptical Gaussian components situated to the north and south of V4641\,Sgr at the 3$\sigma$ confidence level \citep{HESS25}. At the $\gamma$-ray emission regions shown in Figure~\ref{fig: x+gam} and \ref{fig: 4641+433}, the diffuse X-ray structures near V4641\,Sgr, i.e., n1, s1, and s2, look similar to the brightest part of X-ray emission around SS\,433. The n1 shares some similarity with the ``lenticular'' region, comparable to SS\,433 e2 (also see Appendix~\ref{app: mul} for discussion of an alternative scenario, although seemingly not likely). They both have a transverse instead of radial orientation, and both appear near the far-end of the H.E.S.S. source. On the other hand, s1 and s2 fall in the H.E.S.S region, making them somewhat comparable to SS\,433 w1 and w2. 

Nevertheless, we note that the X-ray morphology of diffuse structures of V4641\,Sgr does not perfectly match those around SS\,433. The primary observational difference in X-ray bands between SS\,433 and V4641\,Sgr is the observed flux, although the distance of V4641\,Sgr \citep[$\sim6.2$\,kpc,][]{HESS25} is slightly larger than SS\,433 \citep[$\sim5$\,kpc,][]{Su18}. The luminosity of the diffuse X-rays associated with V4641\,Sgr (i.e., n1, s1, and s2) is $\sim2\times10^{33}$\,erg\,s$^{-1}$, while for SS\,433, the luminosity of the extended jet is $\gtrsim10^{35}$\,erg\,s$^{-1}$ in 0.5--8\,keV \citep{Safi-Harb97, Tong26}. Fainter and more extended emissions may exist near V4641\,Sgr, just like SS\,433, where diffuse X-ray emission nearly fills the whole TeV emitting region. The absence of analogous faint diffuse X-ray emission surrounding V4641\,Sgr could be limited by the sensitivity of EP. Morphologically, as shown in Figure~\ref{fig: 4641+433}, SS\,433 exhibits non-thermal extended X-ray lobes with its near (w1 and e1) and far (w2 and e2) end situated at $\sim25$\,pc and $\sim50$\,pc from the microquasar, respectively. In contrast, the extended structures n1 and s1 of V4641\,Sgr lie merely $\sim23$\,pc from the source. The component s2 resides at $\sim42$\,pc from the microquasar, yet it remains closer than the outermost end of SS\,433's extended $\gamma$-ray emission. Additionally, no analogous bright compact X-ray component corresponding to SS\,433 e1 is detected around V4641\,Sgr. These subtle morphological disparities likely arise from differences in intrinsic properties and surrounding ambient environments of the two microquasars.

\subsection{SED modeling of leptonic emissions}\label{sec:gam}

\begin{figure}
    \centering
    \includegraphics[width=0.47\textwidth]{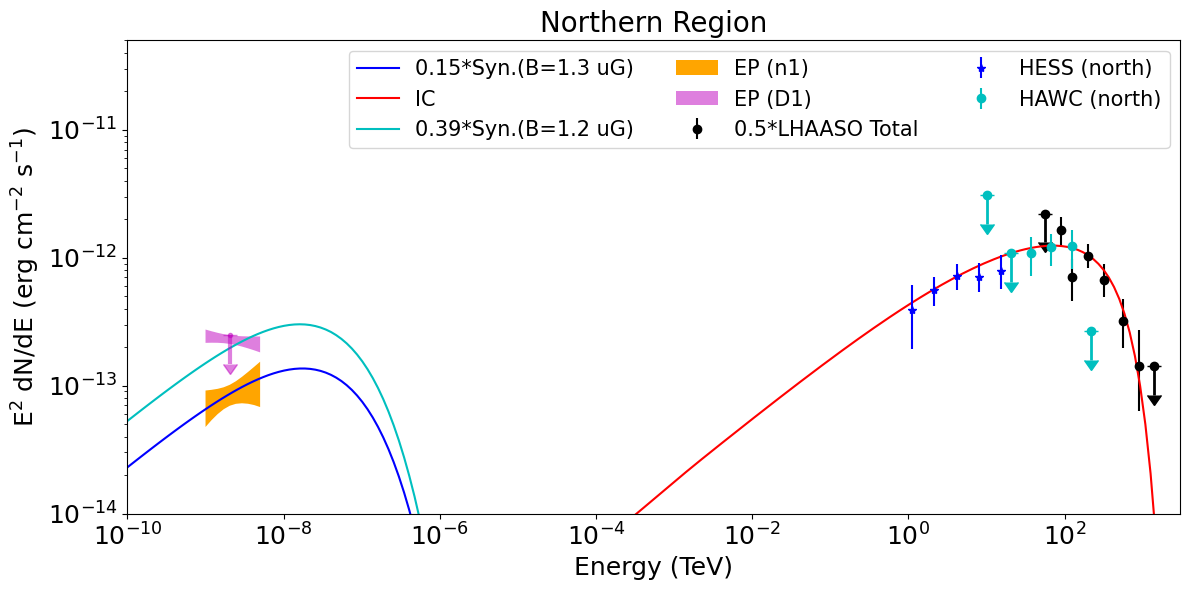}
    \includegraphics[width=0.47\textwidth]{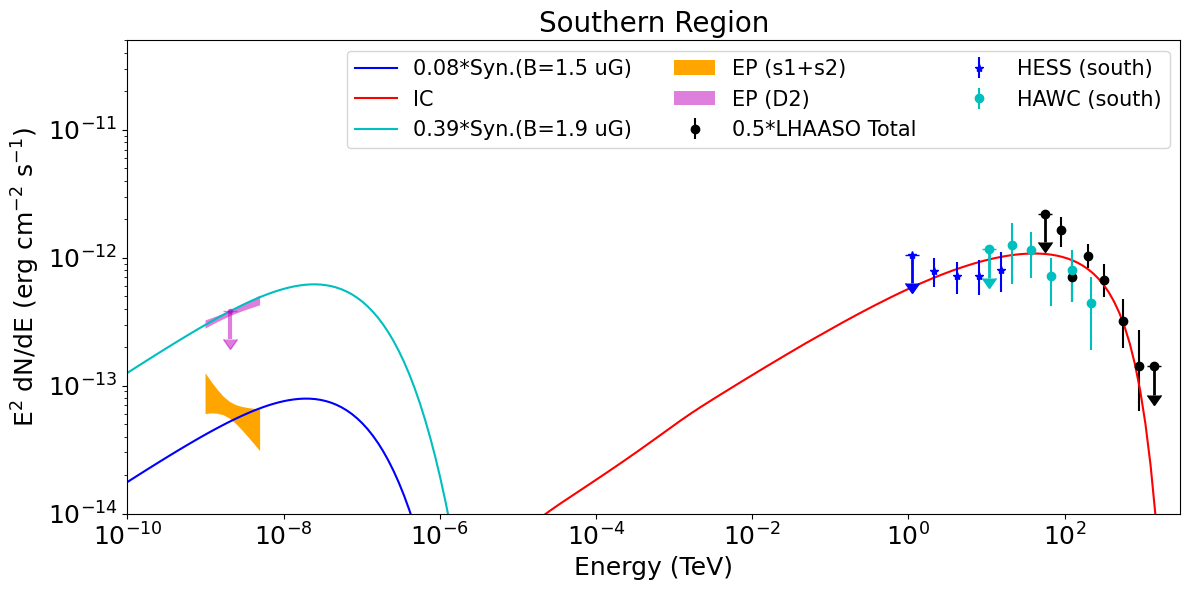}
    \caption{Leptonic emission modeling based on the $\gamma$-ray and X-ray emission measurements. Top panel: multi-wavelength spectrum modeling for the northern region of V4641\,Sgr. The cyan curve represents the synchrotron  flux from the CR electrons inside the D1 region with a magnetic field $B=1.2\,\mu$G. The blue curve represents the synchrotron flux from the electrons inside n1 region with $B=1.3\,\mu$G. Bottom anel: multi-wavelength spectrum modeling for the southern region of V4641\, Sgr. The cyan curve represents the synchrotron flux from the electrons inside the D2 region with $B=1.9\,\mu$G. The blue curve represents the synchrotron flux from the electrons inside s1+s2 region with $B=1.5\,\mu$G.}
    \label{fig: sed}
\end{figure}
Since TeV $\gamma$-ray emission from V4641\,Sgr has been observed by HAWC, LHAASO, and H.E.S.S., several non-thermal emission models have been proposed, including leptonic, hadronic, and lepto-hadronic scenarios \citep{2025arXiv251202574D, 2025PhRvD.111j3025N, 2026ApJ...997..128W, 2026JHEAp..5100538C, 2026PhRvD.113j3053K}. Hadron-dominated models are disfavored due to the energetic problem imposed by the low-density ambient interstellar medium (ISM), as discussed in previous studies \citep{2026ApJ...997..128W,HESS25, 2026PhRvD.113j3053K}. Accordingly, we tentatively adopt a leptonic model to reproduce the multi-wavelength radiation spectrum of V4641\,Sgr.
 
We employ a single-zone model for the northern and southern emission regions separately. Within each zone, X-ray and $\gamma$-ray emissions originate from the same relativistic electron population via synchrotron radiation and inverse Compton (IC) scattering, respectively. Detailed model configurations and flux scaling are presented in Appendix~\ref{app: sed}, and the corresponding SED modeling results are illustrated in Figure~\ref{fig: sed}. 

X-ray measurements allow us to place constraints on the average magnetic field intensity by comparing the X-ray and $\gamma$-ray flux within the regions listed in Table~\ref{tab: xspec}. X-ray flux derived from northern n1 and southern combined s1+s2 regions (listed in Table~\ref{tab: xspec}) constrain the magnetic field to $\sim 1.3\,\mu\mathrm{G}$ and $\sim 1.2\,\mu\mathrm{G}$ in these regions, respectively. Taking a conservative approach, we use the X-ray flux upper limits derived from EP at D1 and D2 to set magnetic field upper bounds of $B<1.2\,\mu\mathrm{G}$ (north) and $B<1.9\,\mu\mathrm{G}$ (south). For nearby areas with lower X-ray surface brightnesses, the magnetic field could be even lower.

Although the flux scaling can affect the estimation of the magnetic field, the estimate is relatively insensitive to this uncertainty, given that $F_\mathrm{X}\propto B^{(\alpha+1)/2}$, where $\alpha\approx2$ is the spectral index of the electron according to the $\gamma$-ray spectra (see Appendix~\ref{app: sed}). For the same reason, a higher magnetic, e.g., $\sim3\,\mu$G, is possible if the $\gamma$-ray emissions are hadronic-dominated, which requires a higher energy budget due to the lack of dense target gas \citep[e.g.,][]{HESS25, 2026PhRvD.113j3053K}. Future neutrino observations are crucial to constrain the fraction of leptonic and hadronic emissions.

On the other hand, low magnetic field was also reported in previous X-ray-to-$\gamma$-ray joint studies. For instance, a low magnetic field of $\sim1.4\,\mu$G was proposed around HESS J1303$-$631 associated with the PWN powered by PSR B1259$-$63 \citep{HESS12} on the Galactic Plane. Actually, a weak magnetic field is not exceptional around V4641\,Sgr considering its location roughly $\sim520$\,pc below the Galactic Plane. At such large vertical displacements from the disk, theoretical Galactic ordered magnetic field models predict typical field intensity of \(\sim1\,\mu\mathrm{G}\) \citep{2019A&A...623A.113S}. 

\subsection{Leptonic emissions around V4641~Sgr}

The low magnetic field implies slower synchrotron cooling, allowing UHE emissions from electrons to be observed. Adopting a magnetic field of 2\,$\mu$G, approximately the upper limit according to our SED fitting, we describe the electron cooling timescale as
\(t_\mathrm{c}\lesssim3.0\,\mathrm{kyr}\,({E_\mathrm{e}}/{1\,\mathrm{PeV}})^{-0.75}\), where $E_\mathrm{e}$ denotes the electron energy. The energy loss due to IC is also considered here, although it is not significant for \(E_\mathrm{e}\sim1\)\,PeV (for details see Appendix~\ref{Sec:cooling}).
At energies above 250\,TeV, LHAASO resolves an extended source with a semi-major axis of up to \(0.86^\circ\) within its 90\% flux region \citep{Cao25}; its emission centroid is located \(\sim0.5^\circ\) from V4641\,Sgr. If the particle acceleration occurs near V4641\,Sgr, we estimate that PeV-scale electrons travel a distance of \(l\sim150\,\text{pc}\).
Under the assumption of diffusive transport, the diffusion coefficient required is \(D_{1\,\text{PeV}}\gtrsim l^2/(4t_\mathrm{c})=6\times10^{29}\,\text{cm}^2\,\text{s}^{-1}\), lower than the typical value in ISM under Kolmogorov diffusion. 

The energy budget poses a challenge when interpreting the $\gamma$-ray emissions around V4641\,Sgr. We estimate the total energy of the electron population responsible for the X-ray to $\gamma$-ray emission as \(W_\mathrm{e}\sim 4.6\times10^{47}\,\mathrm{erg}\). With a cooling timescale of $t_\mathrm{c}\sim3$\,kyr derived above, the leptonic luminosity is roughly $L_\mathrm{e}\sim W_\mathrm{e}/t_\mathrm{c}\sim5\times10^{36}$\,erg\,s$^{-1}$. For comparison, the average X-ray luminosity of V4641\,Sgr has remained low over recent decades \citep[\(\ll 10^{-4}\) of the Eddington luminosity; see][and references therein]{Liu26}. Our long-term monitoring with WXT yields an observed X-ray luminosity of \(\lesssim2.3\times10^{35}\,\mathrm{erg}\,\mathrm{s}^{-1}\) during spring 2026. \citet{Liu26} proposed that V4641\,Sgr may have a large intrinsic luminosity \(L_\mathrm{X}\sim 10^{38}\,\mathrm{erg}\,\mathrm{s}^{-1}\) with substantial obscuration. Alternatively, a low magnetic field allows these UHE electrons to survive for thousands of years. In this sense, if V4641\,Sgr had been more active in the past and injected a large number of UHE electrons, we might still observe their emission today. 

\section{Summary}\label{sec:summary}
In summary, we report the discovery of diffuse X‑ray structures (n1, s1, and s2) around the Galactic microquasar V4641~Sgr, using observations from the Follow‑up X‑ray Telescope onboard the Einstein Probe. These structures are spatially coincident with the TeV-PeV $\gamma$-ray emission detected by LHAASO, HAWC, and H.E.S.S. The weak X-ray emission poses an upper limit of $1-2\,\mu$G for the magnetic field around V4641~Sgr under the one-zone leptonic model. The low magnetic field implies slow cooling of PeV electrons, in favor of the observed UHE $\gamma$-ray emissions. 

\begin{acknowledgments}
We appreciate the valuable comments from the anonymous referee. Y.H.C. and P. Z.\ thank the support by National Natural Science Foundation of China via NSFC 12273010 and the Fundamental Research Funds for the Central Universities No.\ KG202502. R.-Y. L. acknowledges the support of the Basic Research Program of Jiangsu under grant No.~BK20250059. S.Z. C. thanks the support by NFSC 12393851.
This work is based on the data obtained with Einstein Probe, a space mission supported by the Strategic Priority Program on Space Science of Chinese Academy of Sciences, in collaboration with the European Space Agency, the Max-Planck-Institute for extraterrestrial Physics (Germany), and the Centre National d'Études Spatiales (France). This scientific work uses data obtained from Inyarrimanha Ilgari Bundara / the Murchison Radio-astronomy Observatory. We acknowledge the Wajarri Yamaji People as the Traditional Owners and native title holders of the Observatory site. CSIRO's ASKAP radio telescope is part of the Australia Telescope National Facility (https://ror.org/05qajvd42). Operation of ASKAP is funded by the Australian Government with support from the National Collaborative Research Infrastructure Strategy. ASKAP uses the resources of the Pawsey Supercomputing Research Centre. Establishment of ASKAP, Inyarrimanha Ilgari Bundara, the CSIRO Murchison Radio-astronomy Observatory and the Pawsey Supercomputing Research Centre are initiatives of the Australian Government, with support from the Government of Western Australia and the Science and Industry Endowment Fund. This paper includes archived data obtained through the CSIRO ASKAP Science Data Archive, CASDA (https://data.csiro.au). The MeerKAT telescope is operated by the South African Radio Astronomy Observatory, which is a facility of the National Research Foundation, an agency of the Department of Science and Innovation. 
\end{acknowledgments}

\begin{contribution}

Y.H. Chi leads the X-ray analysis and drafts the manuscript supervised by P. Zhou. S.Q. Xi carries out the broadband SED analysis. The observations are conducted under the FXT-LHAASO project coordinated by H. Feng and S.Z. Chen. R.Y. Liu contributes the discussion on the cosmic rays of V4641\,Sgr. The monitor of V4641\,Sgr and the observation schedule are applied by C.C. Jin. N. Tsuji, C. Tresoldi, G. Ponti, K. Mori, and M. Lundy share their XMM-Newton results for the coordination of the publication. All the authors contribute to the improvement of the manuscipt.

\end{contribution}

\facilities{EP(FXT), MeerKAT, ASKAP}

\software{HEASoft, FXTDAS, CIAO
          }
          
\appendix

\section{Footprint of FXT observations}\label{app: obs}

\begin{table*}
    \centering
    \caption{Observation information of V4641\,Sgr. All the observations are applied in the Full-Frame mode with the Thin filter.}
    \begin{tabular}{ccccccccc}
    \hline
    Obs. ID & Obs. Date & R.A. & Decl. & \multicolumn{2}{c}{Exposure} & \multicolumn{2}{c}{Rate$^a$} & COR$^b$ \\
    & & (J2000) & (J2000) & \multicolumn{2}{c}{(ks)} & \multicolumn{2}{c}{(cts\,s$^{-1}$)} & (GV) \\ 
    & & & & FXTA & FXTB  & FXTA & FXTB & \\ \hline
		11900386432 & 2025-09-09 & 274.812 & $-$25.444 & 8.8 & 8.9 & 3.2 & 4.2 & 9.6 \\
		11900399744 & 2025-09-18 & 274.812 & $-$25.444 & 14.4 & 14.8 & 3.2 & 4.2 & 8.8 \\
		11900679680 & 2026-04-10 & 274.839 & $-$25.856 & 1.5 & 1.4 & 3.2 & 4.7 & 8.8 \\
		11900680064 & 2026-04-10 & 274.839 & $-$25.856 & 4.3 & 4.2 & 3.1 & 4.3 & 8.8 \\
		11900684800 & 2026-04-13 & 274.840 & $-$25.856 & 2.7 & 2.7 & 3.2 & 4.5 & 8.8 \\
		06800001354 & 2026-04-13 & 274.839 & $-$25.606 & 4.8 & 5.0 & 3.3 & 4.5 & 8.8 \\
		11900727296 & 2026-05-14 & 274.881 & $-$25.586 & 2.1 & 2.2 & 3.4 & 4.3 & 8.7 \\
		11900731520 & 2026-05-17 & 274.881 & $-$25.587 & 3.5 & 3.4 & 3.0 & 3.8 & 10.2 \\
		11900733056 & 2026-05-20 & 274.881 & $-$25.587 & 3.0 & 3.0 & 3.4 & 3.9 & 8.8 \\
		11900734592 & 2026-05-21 & 274.881 & $-$25.587 & 3.5 & 3.5 & 3.7 & 4.3 & 8.9 \\
		11900743424 & 2026-05-27 & 274.882 & $-$25.587 & 1.8 & 1.7 & 2.9 & 3.7 & 9.0 \\
		11900747008 & 2026-05-30 & 274.882 & $-$25.587 & 2.7 & 2.8 & 3.0 & 3.6 & 8.8 \\
		11900751744 & 2026-06-03 & 274.883 & $-$25.587 & 1.7 & 1.8 & 3.1 & 3.7 & 8.8 \\
		11900755328 & 2026-06-06 & 274.884 & $-$25.587 & 2.0 & 2.0 & 2.9 & 3.5 & 8.8 \\
		11900756352 & 2026-06-07 & 274.884 & $-$25.587 & 1.0 & 0.8 & 3.0 & 3.5 & 8.8 \\
		11900763392 & 2026-06-12 & 274.886 & $-$25.589 & 3.9 & 4.1 & 3.0 & 3.5 & 8.8 \\
        11900763776 & 2026-06-13 & 274.885 & $-$25.589 & 2.6 & 2.5 & 2.9 & 3.5 & 10.1 \\ \hline
        Total & & & & 64.3 & 64.8 & & \\ \hline
    \end{tabular}
    \begin{tablenotes}
        \item {\bf Notes:} $^a$ In 0.5--7\,keV band. $^b$ Cut off rigidity.
    \end{tablenotes}
    \label{tab: obsinfo}
\end{table*}

\begin{figure*}
    \centering
    \includegraphics[width=0.8\textwidth]{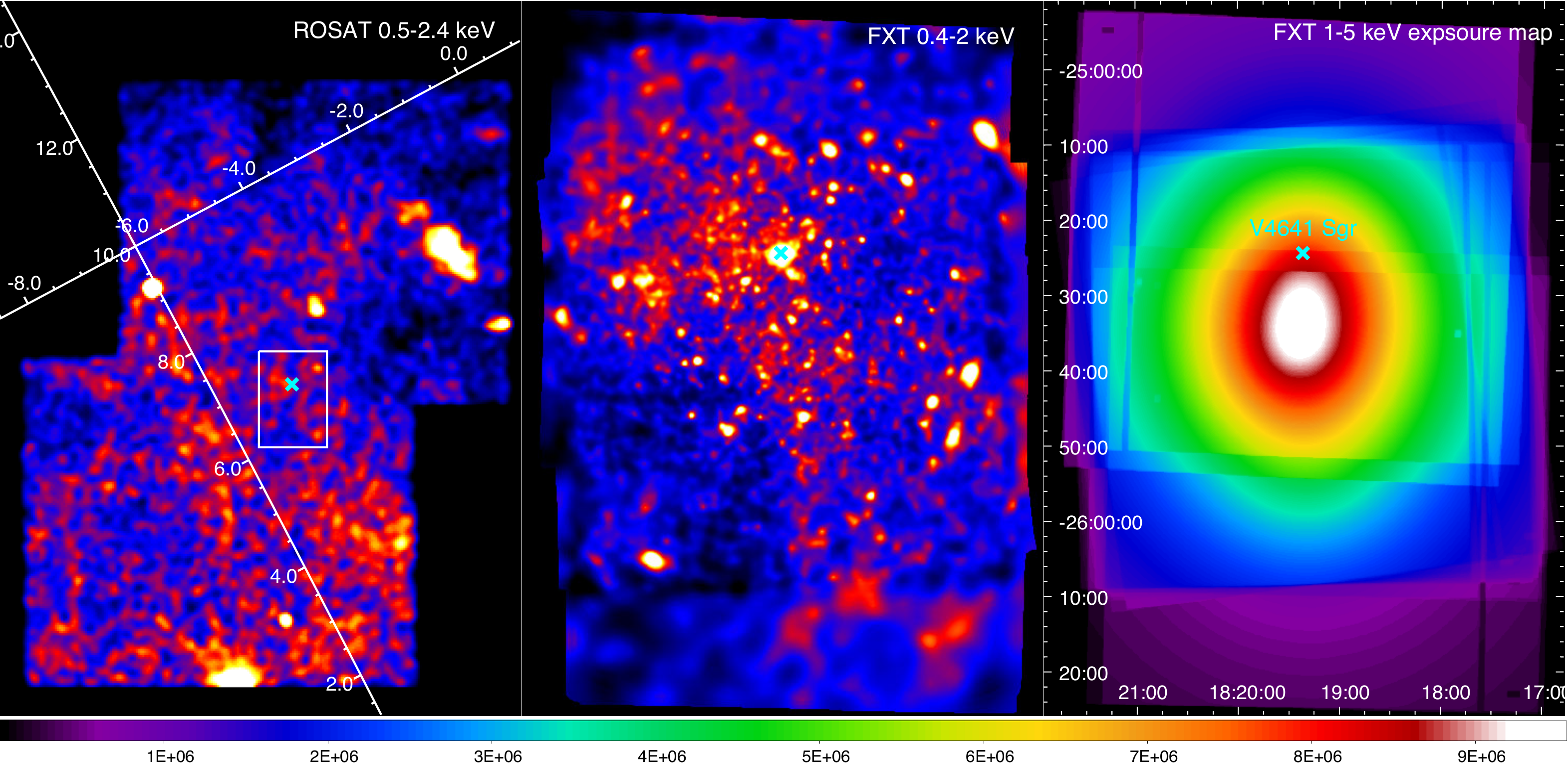}
    \caption{Left: ROSAT all-sky survey rate map of the region around V4641\,Sgr in 0.5--2.4\,keV. The white box marks the coverage of EP-FXT. Middle: adaptively smoothed vignetting-corrected FXT image in 0.4--2\,keV. Right and colorbar: the exposure map of FXT in 1--5\,keV band in units of s\,cm$^{2}$. V4641\,Sgr is labled with a cyan cross.}
    \label{fig: rass}
\end{figure*}

The FXT observations consist of 17 individual epochs with a total exposure of $\sim64$\,ks. The detailed information is listed in Table~\ref{tab: obsinfo}. We found that the average count rate of the whole field of view did not show evident variation among different observations. The cut-off rigidity of the observations recorded in the MKF files is $\gtrsim8$\,GV, indicating that the contamination from geomagnetic activities are negligible.

Two observations ($\sim22$\,ks in total) aiming at V4641\,Sgr were performed in the autumn of 2025. The others were applied in spring of 2026, focusing on the regions south of V4641\,Sgr. Hence, only the 2025 observations were used to analyze the spectra of n1.

The footprint of FXT observations is also shown in Figure~\ref{fig: rass} with a ROSAT image. The ROSAT image consists of two $3^\circ\times3^\circ$ tiles of the all-sky survey, rs932245n00 and rs932147n00. We retrieved the 0.5--2.4\,keV counts maps and exposure maps of individual observations from the HEASARC Browse\footnote{https://heasarc.gsfc.nasa.gov/cgi-bin/W3Browse/w3browse.pl}, which were later mosaicked with the CIAO script {\it reproject\_image\_grid}. The mosaicked counts map was divided by the exposure map to generate the rate map. No evident structure (e.g., a bubble) is detected around V4641\,Sgr, and it is consistent with the FXT image, where the southeastern part looks fainter.

\section{Sky modeling}\label{app: bkg}

We modeled the sky background to handle the response among different regions. We considered an XSPEC model {\it apec}+{\it TBabs}*({\it apec}+{\it powerlaw}), motivated by some astrophysical components (see, e.g., XMM-ESAS cookbook\footnote{https://heasarc.gsfc.nasa.gov/docs/xmm/esas/cookbook/xmm-esas.html}) including (1) an unabsorbed plasma component with $kT\sim0.1$\,keV representing local hot gas like the local hot bubble, (2) an absorbed plasma component with $kT\sim0.4\,$keV standing for the Galactic hot gas, and (3) a non-thermal component with its photon index fixed to 1.45 from unresolved sources like background active galaxy nuclei. We let the temperatures, absorption, and fluxes of these components vary in the fitting for a better mimic. As the background model is simply fixed and scaled in the fitting of the off-source spectra, the actual parameters and their interpretations are not crucial in our studies.

For the H.E.S.S. regions (i.e., D1 and D2), as there are no regions for a sky background, we used a similar model as mentioned above, where the non-thermal component consists of both sky background and potential synchrotron emissions associated with UHE sources. The column density was fixed to $2\times10^{21}$\,cm$^{-2}$ to be consistent with the analysis of n1, s1, and s2. 

\begin{figure}
    \centering
    \subfigure{\includegraphics[width=0.49\textwidth]{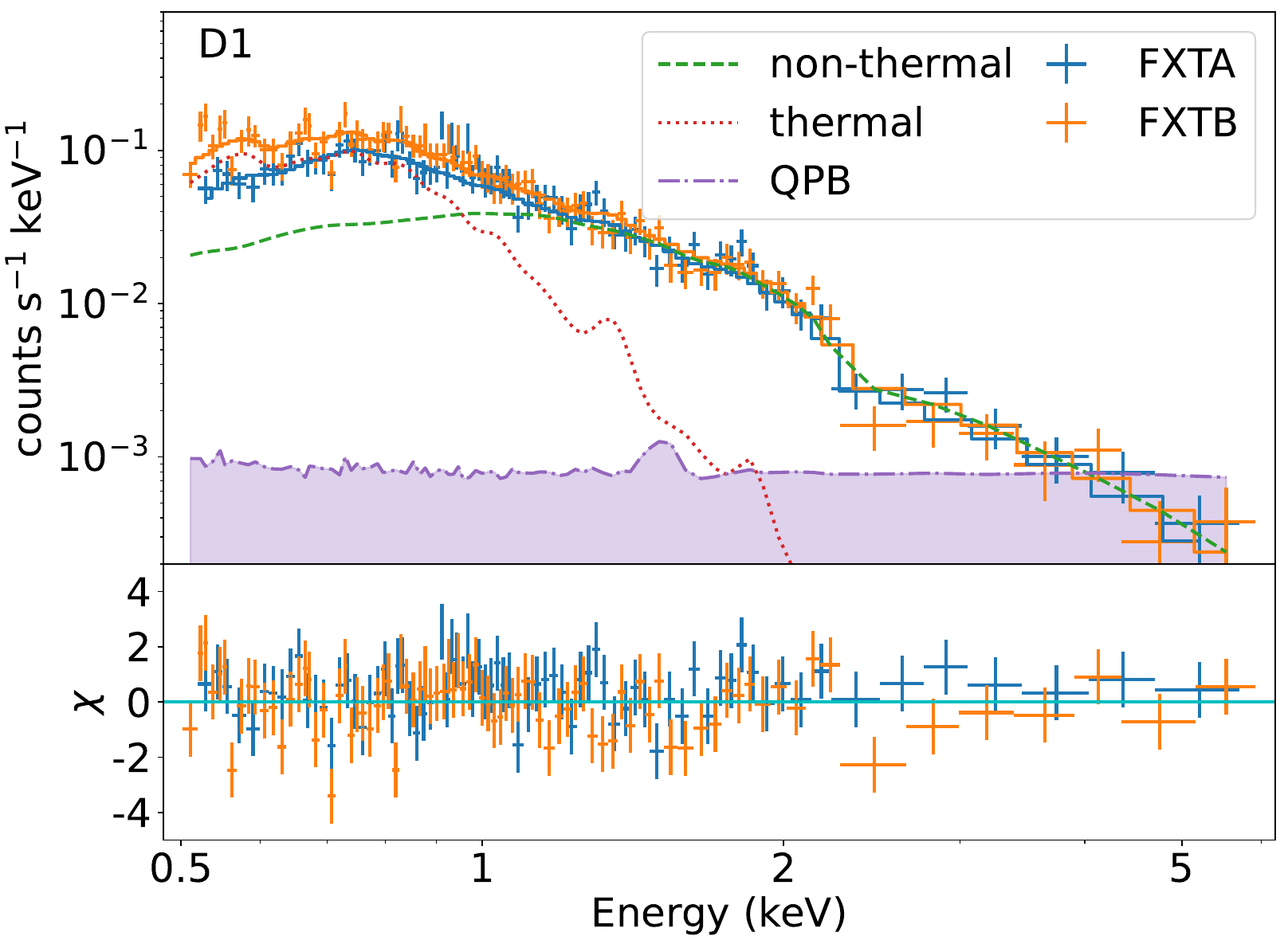}}
    \subfigure{\includegraphics[width=0.49\textwidth]{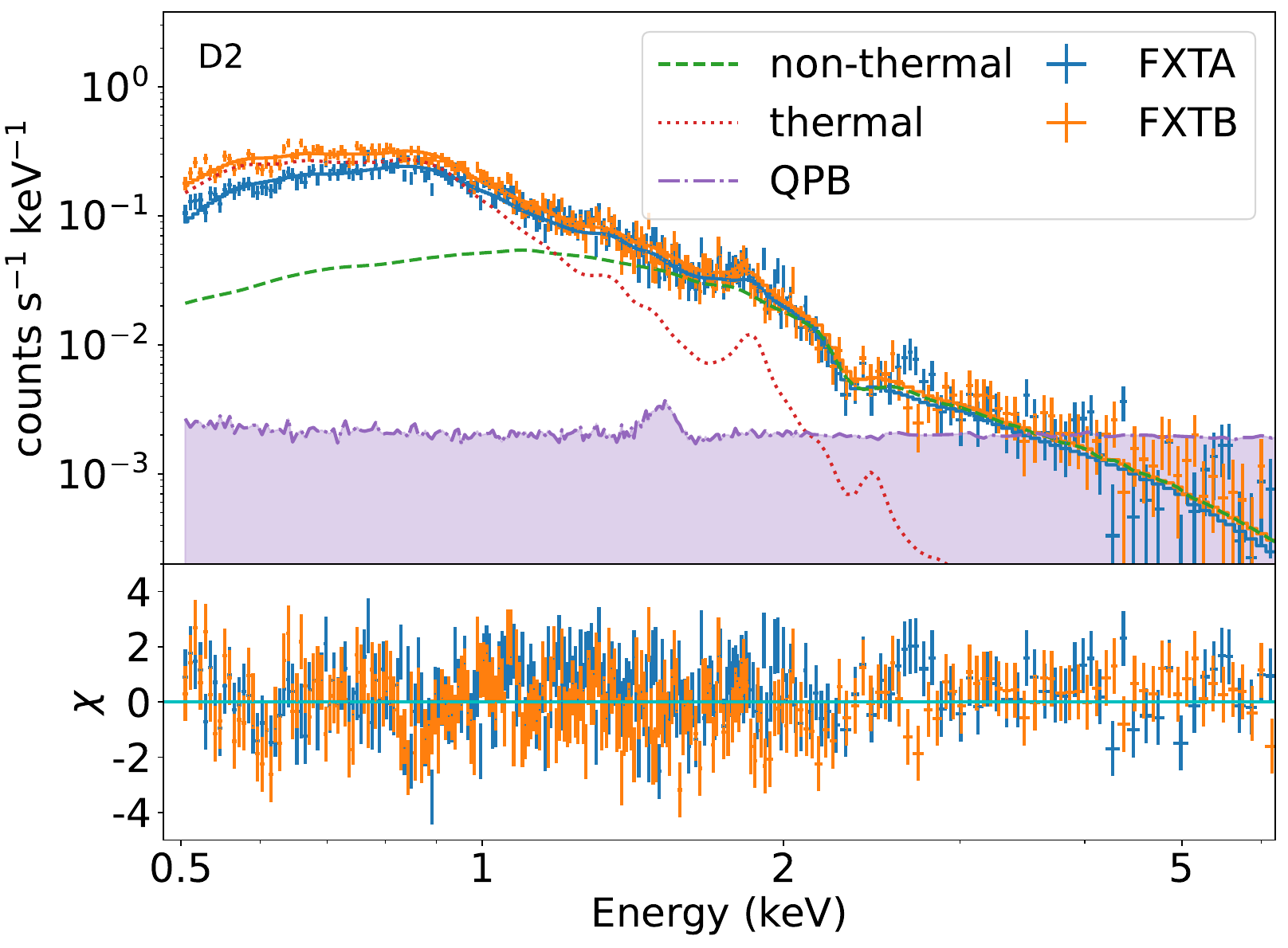}}
    \caption{Fitted spectra with folded models of D1 (upper panel) and D2 (lower panel). The purple lines mark the instrumental background level, which is subtracted from the observed spectra. The dashed green lines are non-thermal components (including sky background) while the dotted red lines are thermal emissions (e.g., foreground plasma).}
    \label{fig: dspec}
\end{figure}

Figure~\ref{fig: dspec} shows the fitted spectra with folded models of region D1 and D2. Different components are labeled as well.

\section{The point-like source in n1}\label{app: mul}
\begin{figure}
    \centering
    \includegraphics[width=0.45\textwidth]{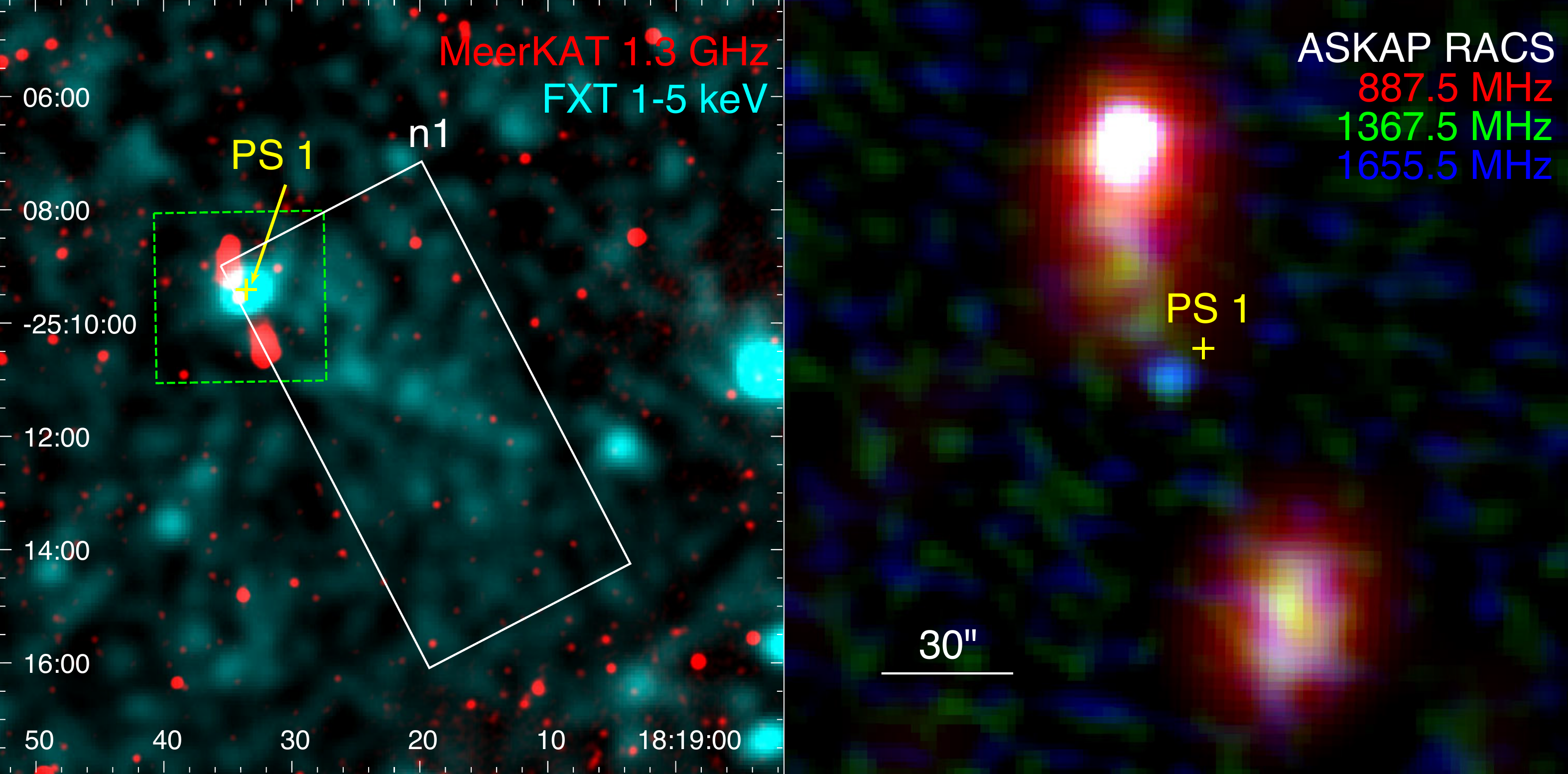}
    \caption{Left panel: MeerKAT 1.3\,GHz observation of n1 (red) overlapped by the FXT 1-5\,keV image smoothed with a kernel size of $24''$. Right panel: Zoomed-in ASKAP RACS survey (red/green/blue: 887.5/1367.5/1655.5\,MHz) of the green region in the left panel. The position of the X-ray point-like source PS\,1 is labeled in yellow.}
    \label{fig: ps1}
\end{figure}

A point-like X-ray source appears at the left end of n1, named ``PS\,1'' hereafter (see Figure~\ref{fig: ps1}). Limited to its low count rate ($\sim120$ net counts derived), a search for pulses is unfeasible but its flux is comparable between two epochs with a rate of $\sim5.6\times10^{-3}$\,counts\,s$^{-1}$ for two FXT modules. The background spectra of PS\,1 were extracted from an annulus region and directly subtracted from the source spectra. 

\begin{table}
    \centering
    \caption{Fitting results of PS\,1. The foreground absorption is fixed at $2\times10^{21}$\,cm$^{-2}$.}
    \begin{tabular}{ccccc}
    \hline
        Model & $kT$ (keV) & $\Gamma$ & $F^a$ & $\chi^2$/dof \\ \hline
        BB & $0.69_{-0.14}^{+0.19}$ & $\cdots$ & $9.3_{-2.8}^{+3.4}$ & 17.75/20 \\
        PL & $\cdots$ & $0.77\pm0.34$ & $23.6\pm7.6$ & 17.31/20 \\ \hline
    \end{tabular}
    \begin{tablenotes}
        \item {\bf Notes:}$^a$ Flux in 0.5--8\,keV in units of $10^{-14}$\,erg\,s$^{-1}$\,cm$^{-2}$.
    \end{tablenotes}
    \label{tab: ps1}
\end{table}

Motivated by the belt-like morphology of n1 and PS\,1's position at the northeastern end, one possible explanation of the X-ray emission is a pulsar with a tail/nebula of synchrotron X-ray emissions. 

We fit the spectra of PS\, 1 using a blackbody (BB) model and a power-law (PL) model, with the foreground absorption fixed to $2\times 10^{21}\, \rm cm^{-2}$ (see Table~\ref{tab: ps1}).
However, the X-ray spectral analysis does not strongly support PS\,1 to be a pulsar. The thermal BB model of the X-ray emission gives a relatively high temperature of $\sim0.6$\,keV and a small emission size of around 0.16$D_{6.2}$\,km. 
For isolated pulsars, the hottest ones with temperatures of $\sim0.4$--0.5\,keV are very young pulsars in SNRs and their emission sizes would be $\gtrsim1$\,km \citep[see][and references therein]{Yu11}. 
For the non-thermal PL model, the photon index of $0.77\pm 0.34$ seems too small \citep{Borghese26} compared to that in pulsars.

We also searched for potential radio counterparts of PS\, 1 and n1, using MeerKAT \citep{Noa26} and ASKAP \citep{askap1, askap2, askap5, askap6} data as shown in Figure~\ref{fig: ps1}. No diffuse radio emissions are observed associated with n1, but one point-like radio source is shown within the FXT's astrometry uncertainty of $\sim7''$ from PS\,1. To the north and south of this radio source, two lobe-like structures are evident, making these emissions look like radio jets from an active galactic nucleus (AGN). The two lobes show a negative radio index (i.e., brighter in the 887.5\,MHz band), while the index of the central point-like source is positive. A negative index is expected for the non-thermal emission from a pulsar except for fierce low-frequency absorption. Therefore, there lacks solid evidence for a common origin of PS\,1 and n1.

\section{Details of the SED fitting}\label{app: sed}
In the pure leptonic scenario, the X-ray and $\gamma$-ray emissions are produced via synchrotron radiation and inverse Compton (IC) scattering, respectively, from a single electron population. We assume the steady-state electron energy spectrum follows a power-law distribution with a super-exponential cutoff:
\begin{equation}\label{splc}
    \frac{dN_{\rm e}}{dE_{\rm e}} \propto E_{\rm e}^{-\alpha} \exp\left[-\left(\frac{E_{\rm e}}{E_{\rm c}}\right)^2\right] \, ,
\end{equation}
where $E_{\rm e}$ is the primary particle energy, $\alpha$ is the spectral index, and $E_{\rm c}$ is the exponential cutoff energy. The value of the parameters are listed in Table~\ref{tab: sedpar}. The photon field contains the Cosmic Microwave Background (CMB) along with a far-infrared (FIR) emission component. The FIR emission corresponds to a gray-body spectrum of $T_{\rm FIR} = 30$\,K and energy density $u_{\rm FIR} = 0.5\,\rm eV\,cm^{-3}$ according to an interstellar radiation field model \cite{Popescu2017}. The electron energy spectrum can be tightly constrained through measurements of the $\gamma$-ray spectrum. The two-source scenario is considered here, which is supported with a significance of $\sim3\sigma$ according to the H.E.S.S. observations \citep{HESS25}. The spectra of H.E.S.S. and HAWC are derived from \citet{HESS25} and \citet{HAWC24}, respectively. For LHAASO, as two sources share high similarities according to H.E.S.S. and HAWC, we simply supposed that each source contributes half the flux of the total LHAASO spectrum \citep{Cao25}.

\begin{table}
    \centering
    \caption{Parameters of the Leptonic model}
    \begin{tabular}{cccc}
    \hline
        Parameter & Unit & North & South \\ \hline
        $\alpha$ &  & 2.0 & 2.2 \\
        $E_\mathrm{c}$ & TeV & 800 & 900 \\
        $W_\mathrm{e,>1 GeV}$ & \(10^{47}\)\,erg & 1 & 3.6 \\ \hline
    \end{tabular}
    \label{tab: sedpar}
\end{table}

Among existing TeV observations, H.E.S.S. provides the best spatial resolution, supporting two elliptical Gaussian structures located to the north and south of the microquasar V4641\,Sgr. We therefore adopt the H.E.S.S.-resolved sources D1 and D2 to characterize the $\gamma$-ray morphology of V4641\,Sgr across TeV to PeV energies. For each given X-ray region, we calculated the encircled IC flux fraction according to the H.E.S.S. 2-D elliptical Gaussian morphology, which is then used to estimate the average magnetic field with X-ray-to-$\gamma$-ray SED fitting in the X-ray region. For instance, D1 and D2 encircle 39\% of the total $\gamma$-ray flux corresponding to the 1$\sigma$ extension of the Gaussian distribution. Therefore, the $\gamma$-ray flux is scaled to 39\% to estimate magnetic field in these two regions. The regions n1 and s1+s2 correspond to 15\% and 8\% of the total $\gamma$-ray flux of the northern and southern $\gamma$-ray sources, respectively.

The $\gamma$-ray data from H.E.S.S., HAWC, and LHAASO span a broad energy range from $\sim1$\,TeV to $\sim1$\,PeV. The energies of the parent electrons are expected to be comparable to or slightly higher than these $\gamma$-ray energies. For given magnetic field and electron energy, the characteristic energy of synchrotron photons is $E_\mathrm{crit}=(1.3\,\mathrm{keV})(E_\mathrm{e}/0.1\,\mathrm{PeV})^2(B/2\,\mu\mathrm{G})$. Therefore, the IC emission from these X-ray-emitting electrons falls within the energy range covered by the $\gamma$-ray data considered in this work.

\section{The minimal cooling time estimation} \label{Sec:cooling} 

\begin{figure}[ht]
    \centering
    \centering
    \includegraphics[width=0.45\textwidth]{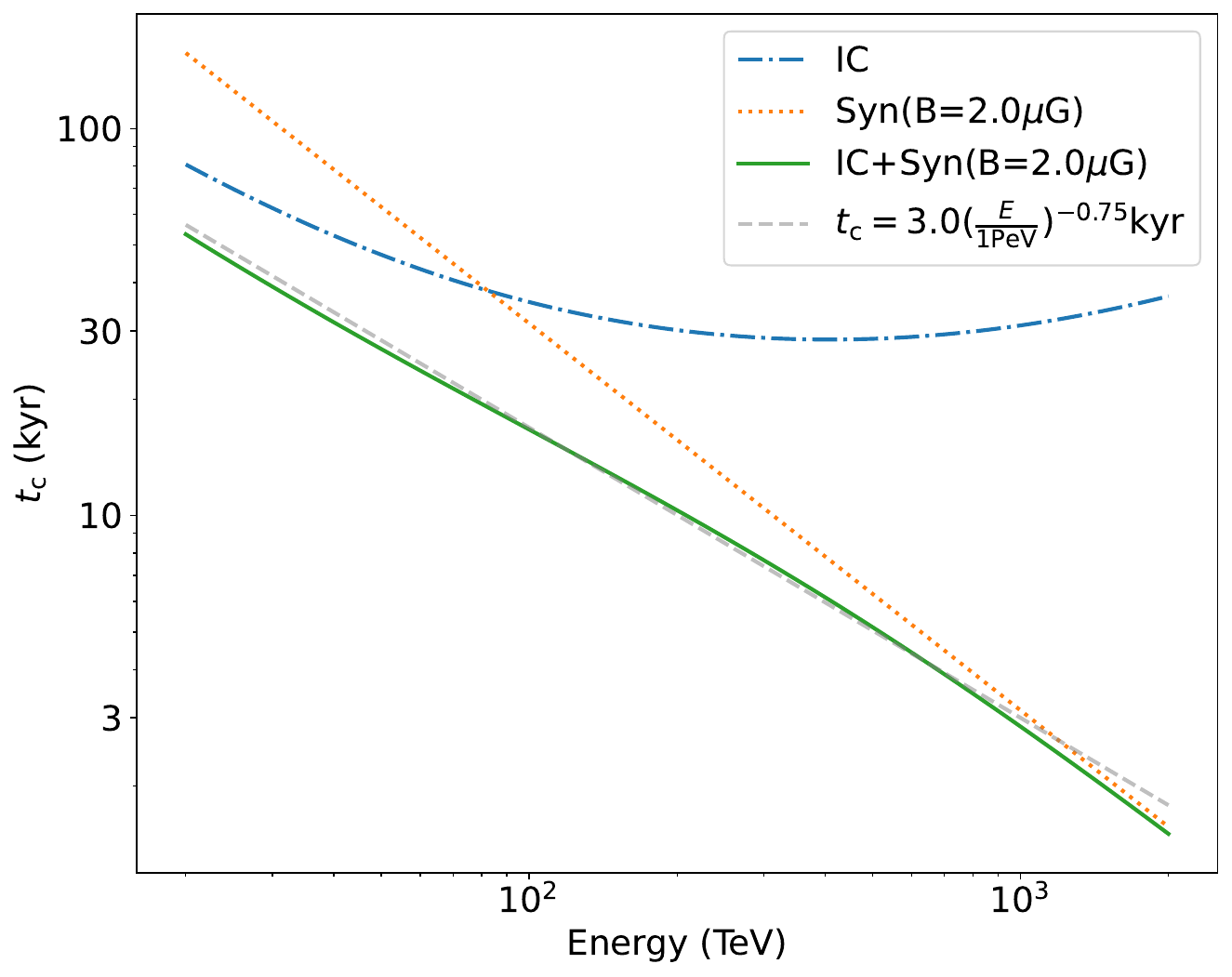} \\
    \includegraphics[width=0.45\textwidth]{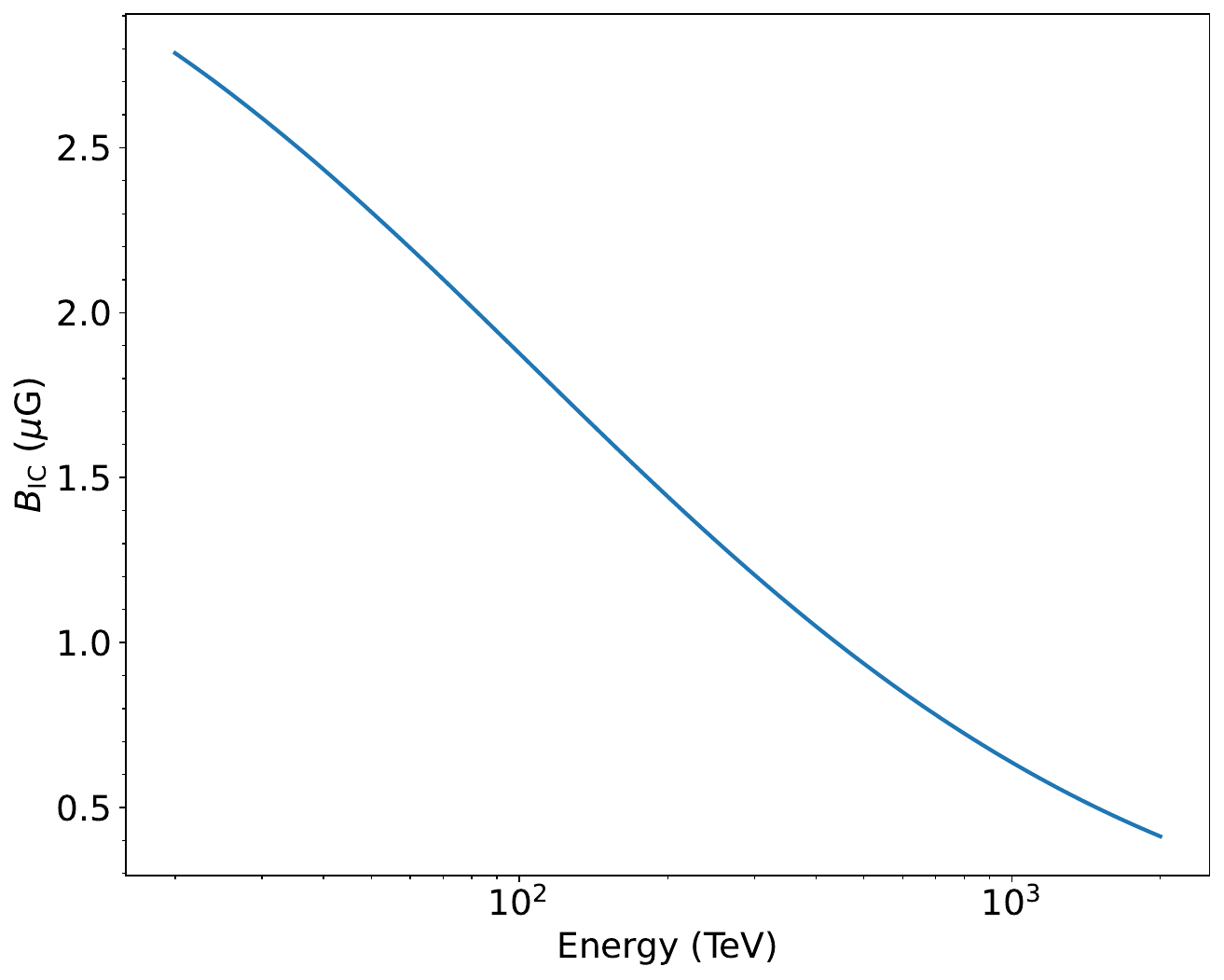}
    \caption{Upper panel: the cooling timescale of electrons versus the electron energy due to IC (blue dash-dotted line) and synchrotron (orange dotted line), respectively. The cooling curve due to IC and synchrotron is labeled with green solid line and mimicked by a powerlaw function (grey dashed line). See also \citet{HESS25} for a similar plot for 3\,$\mu$G field. Lower panel: the effective magnetic field due to IC, which decreases as the energy increases due to the Klein-Nishina limit.}
    \label{fig:coolingtime}
\end{figure}

Here, we investigate the electron cooling time as a function of electron energy using the open-source package GAMERA\footnote{https://libgamera.github.io/GAMERA/docs/tutorials\_main.html}, incorporating both synchrotron and inverse-Compton (IC) energy losses.
For IC scattering, we include the CMB and FIR photon fields, consistent with those adopted in the preceding SED fitting procedure.
Our SED fitting constrains the magnetic field strength to be less than \(2\,\mu\text{G}\). Assuming a uniform magnetic field of \(2\,\mu\text{G}\) across the entire emission region of V4641\,Sgr, we compute the combined synchrotron+IC cooling time as a function of electron energy, which is mimicked with a powerlaw function. The results are presented in Figure~\ref{fig:coolingtime}.
Since the adopted magnetic field corresponds to the upper limit allowed by the SED fitting in one-zone approximation, the cooling times shown in this figure represent the minimum cooling times achievable within the source region of V4641\,Sgr.

On the other hand, we also calculated the ``effective magnetic field'' of IC, defined as
\begin{equation}
    \frac{B_\mathrm{IC}}{B}=\left(\frac{t_\mathrm{c,IC}}{t_\mathrm{c,syn}}\right)^{-\frac{1}{2}},
\end{equation}
where $t_\mathrm{c,IC}$ and $t_\mathrm{c,syn}$ are cooling timescales due to IC and synchrotron, respectively. As $B_\mathrm{IC}\approx0.5\,\mu\mathrm{G}$ near 1\,PeV, IC contributes little to the total effective magnetic field $B_\mathrm{eff}=\sqrt{B^2+B_\mathrm{IC}^2}$.

\bibliography{sample701}{}
\bibliographystyle{aasjournalv7}

\end{CJK*}
\end{document}